\documentclass[10pt]{IEEEtran}
\usepackage{url}
\usepackage[utf8]{inputenc}
\usepackage{color}
\usepackage{amsmath}
\usepackage{amssymb}
\usepackage{cite}

\usepackage[acronyms,nonumberlist,nopostdot,nomain,nogroupskip]{glossaries}
\usepackage{tablefootnote}
\usepackage{booktabs}

\usepackage{tikz}
\usepackage{pgfplots}
\pgfplotsset{compat=newest} 
\pgfplotsset{plot coordinates/math parser=false} 
\newlength\fheight
\newlength\fwidth
\usetikzlibrary{plotmarks,patterns,decorations.pathreplacing,backgrounds,calc,arrows,arrows.meta,spy,matrix}
\usepgfplotslibrary{patchplots,groupplots}
\usepackage{tikzscale}
\usepackage{hyperref}

\usepackage{multirow}
\usepackage{tkz-kiviat}

\usepackage{tabularx}

\usepackage[font=scriptsize]{subcaption}
\usepackage[font=small]{caption}
\usepackage{float}

\usepackage{colortbl}

\usepackage{mathtools}

\newacronym{3gpp}{3GPP}{3rd Generation Partnership Project}
\newacronym{adc}{ADC}{Analog to Digital Converter}
\newacronym{ar}{AR}{Augmented Reality}
\newacronym{vr}{VR}{Virtual Reality}
\newacronym{5g}{5G}{5th generation}
\newacronym{aimd}{AIMD}{Additive Increase Multiplicative Decrease}
\newacronym{am}{AM}{Acknowledged Mode}
\newacronym{amc}{AMC}{Adaptive Modulation and Coding}
\newacronym{aqm}{AQM}{Active Queue Management}
\newacronym{awgn}{AGWN}{Additive White Gaussian Noise}
\newacronym{balia}{BALIA}{Balanced Link Adaptation}
\newacronym{bdp}{BDP}{Bandwidth-Delay Product}
\newacronym{cps}{CPS}{Cyber Physical Systems}
\newacronym{bf}{BF}{Beamforming}
\newacronym{cc}{CC}{Congestion Control}
\newacronym{cdf}{CDF}{Cumulative Distribution Function}
\newacronym{cn}{CN}{Core Network}
\newacronym{cqi}{CQI}{Channel Quality Information}
\newacronym{cp}{CP}{Control Plane}
\newacronym{csirs}{CSI-RS}{Channel State Information - Reference Signal}
\newacronym{dc}{DC}{Dual Connectivity}
\newacronym{umts}{UMTS}{Universal Mobile Telecommunication Systems}
\newacronym{dce}{DCE}{Direct Code Execution}
\newacronym{dci}{DCI}{Downlink Control Information}
\newacronym{dl}{DL}{Downlink}
\newacronym{dmr}{DMR}{Deadline Miss Ratio}
\newacronym{dmrs}{DMRS}{DeModulation Reference Signal}
\newacronym{e2e}{E2E}{End-to-End}
\newacronym{ecn}{ECN}{Explicit Congestion Notification}
\newacronym{edf}{EDF}{Earliest Deadline First}
\newacronym{enb}{eNB}{evolved Node Base}
\newacronym{epc}{EPC}{Evolved Packet Core}
\newacronym{es}{ES}{Edge Server}
\newacronym{fdma}{FDMA}{Frequency Division Multiple Access}
\newacronym{fdd}{FDD}{Frequency Division Duplexing}
\newacronym[firstplural=Radio Access Technologies (RATs)]{rat}{RAT}{Radio Access Technology}
\newacronym{fs}{FS}{Fast Switching}
\newacronym{ftp}{FTP}{File Transfer Protocol}
\newacronym{gnb}{gNB}{Next Generation Node Base}
\newacronym{harq}{HARQ}{Hybrid Automatic Repeat reQuest}
\newacronym{hetnet}{HetNet}{Heterogeneous Network}
\newacronym{hh}{HH}{Hard Handover}
\newacronym{hol}{HOL}{Head-of-Line}
\newacronym{ia}{IA}{Initial Access}
\newacronym{imt}{IMT}{International Mobile Telecommunication}
\newacronym{iot}{IoT}{Internet of Things}
\newacronym{los}{LOS}{Line-of-Sight}
\newacronym{lte}{LTE}{Long Term Evolution}
\newacronym{m2m}{M2M}{Machine to Machine}
\newacronym{mac}{MAC}{Medium Access Control}
\newacronym{mc}{MC}{Multi-Connectivity}
\newacronym{mcs}{MCS}{Modulation and Coding Scheme}
\newacronym{mec}{MEC}{Mobile Edge Cloud}
\newacronym{mi}{MI}{Mutual Information}
\newacronym{mimo}{MIMO}{Multiple Input, Multiple Output}
\newacronym{mmwave}{mmWave}{millimeter wave}
\newacronym{mptcp}{MPTCP}{Multipath TCP}
\newacronym{mr}{MR}{Maximum Rate}
\newacronym{mss}{MSS}{Maximum Segment Size}
\newacronym{mtd}{MTD}{Machine-Type Device}
\newacronym{mtu}{MTU}{Maximum Transmission Unit}
\newacronym{nfv}{NFV}{Network Function Virtualization}
\newacronym{nlos}{NLOS}{Non-Line-of-Sight}
\newacronym{nr}{NR}{New Radio}
\newacronym{ofdm}{OFDM}{Orthogonal Frequency Division Multiplexing}
\newacronym{pdcch}{PDCCH}{Physical Downlink Control Channel}
\newacronym{pdcp}{PDCP}{Packet Data Convergence Protocol}
\newacronym{pdsch}{PDSCH}{Physical Downlink Shared Channel}
\newacronym{pdu}{PDU}{Packet Data Unit}
\newacronym{pf}{PF}{Proportional Fair}
\newacronym{pgw}{PGW}{Packet Gateway}
\newacronym{phy}{PHY}{Physical}
\newacronym{pbch}{PBCH}{Physical Broadcast Channel}
\newacronym[plural=\gls{mme}s,firstplural=Mobility Management Entities (MMEs)]{mme}{MME}{Mobility Management Entity}
\newacronym{prb}{PRB}{Physical Resource Block}
\newacronym{pss}{PSS}{Primary Synchronization Signal}
\newacronym{pucch}{PUCCH}{Physical Uplink Control Channel}
\newacronym{pusch}{PUSCH}{Physical Uplink Shared Channel}
\newacronym{rach}{RACH}{Random Access Channel}
\newacronym{ran}{RAN}{Radio Access Network}
\newacronym{red}{RED}{Random Early Detection}
\newacronym{rf}{RF}{Radio Frequency}
\newacronym{rlc}{RLC}{Radio Link Control}
\newacronym{rlf}{RLF}{Radio Link Failure}
\newacronym{rrc}{RRC}{Radio Resource Control}
\newacronym{rrm}{RRM}{Radio Resource Management}
\newacronym{rr}{RR}{Round Robin}
\newacronym{rs}{RS}{Remote Server}
\newacronym{rsrp}{RSRP}{Reference Signal Received Power}
\newacronym{rss}{RSS}{Received Signal Strength}
\newacronym{rtt}{RTT}{Round Trip Time}
\newacronym{rw}{RW}{Receive Window}
\newacronym{rx}{RX}{Receiver}
\newacronym{sa}{SA}{standalone}
\newacronym{sack}{SACK}{Selective Acknowledgment}
\newacronym{sap}{SAP}{Service Access Point}
\newacronym{sch}{SCH}{Secondary Cell Handover}
\newacronym{scoot}{SCOOT}{Split Cycle Offset Optimization Technique}
\newacronym{sdma}{SDMA}{Spatial Division Multiple Access}
\newacronym{sinr}{SINR}{Signal to Interference plus Noise Ratio}
\newacronym{sm}{SM}{Saturation Mode}
\newacronym{snr}{SNR}{Signal-to-Noise-Ratio}
\newacronym{son}{SON}{Self-Organizing Network}
\newacronym{ss}{SS}{Synchronization Signal}
\newacronym{srs}{SRS}{Sounding Reference Signal}
\newacronym{sss}{SSS}{Secondary Synchronization Signal}
\newacronym{tb}{TB}{Transport Block}
\newacronym{tcp}{TCP}{Transmission Control Protocol}
\newacronym{tdd}{TDD}{Time Division Duplexing}
\newacronym{tdma}{TDMA}{Time Division Multiple Access}
\newacronym{tfl}{TfL}{Transport for London}
\newacronym{tm}{TM}{Transparent Mode}
\newacronym{trp}{TRP}{Transmitter Receiver Pair}
\newacronym{tti}{TTI}{Transmission Time Interval}
\newacronym{ttt}{TTT}{Time-to-Trigger}
\newacronym{tx}{TX}{Transmitter}
\newacronym{ue}{UE}{User Equipment}
\newacronym{ul}{UL}{Uplink}
\newacronym{uml}{UML}{Unified Modeling Language}
\newacronym{um}{UM}{Unacknowledged Mode}
\newacronym{utc}{UTC}{Urban Traffic Control}
\newacronym{vm}{VM}{Virtual Machine}
\newacronym{rsrq}{RSRQ}{Reference Signal Received Quality}
\newacronym{rssi}{RSSI}{Received Signal Strength Indicator}
\newacronym{crs}{CRS}{Cell Reference Signal}
\newacronym{nsa}{NSA}{Non Stand Alone}
\newacronym{mrdc}{MR-DC}{Multi \gls{rat} \gls{dc}}
\newacronym{endc}{EN-DC}{E-UTRAN-\gls{nr} \gls{dc}}
\newacronym{5gc}{5GC}{5G Core}
\newacronym{si}{SI}{Study Item}
\newacronym{iab}{IAB}{Integrated Access and Backhaul}
\newacronym{wf}{WF}{Wired-first}
\newacronym{hqf}{HQF}{Highest-quality-first}
\newacronym{pa}{PA}{Position-aware}
\newacronym{mlr}{MLR}{Maximum-local-rate}
\newacronym{wbf}{WBF}{Wired Bias Function}
\newacronym{mib}{MIB}{Master Information Block}
\newacronym{sib}{SIB}{Secondary Information Block}
\newacronym{kpi}{KPI}{Key Performance Indicator}
\newacronym{ppp}{PPP}{Poisson Point Process}
\newacronym{gtp}{GTP}{GPRS Tunneling Protocol}
\newacronym{amf}{AMF}{Access and Mobility Management Function}
\newacronym{vlc}{VLC}{Visible Light Communications}
\newacronym{led}{LED}{Light Emitting Diode}
\newacronym{qos}{QoS}{Quality of Service}
\newacronym{das}{DAS}{Distributed Antenna System}
\newacronym{mtc}{MTC}{machine-type communications}
\newacronym{ml}{ML}{machine learning}

\makeglossaries
\begin{document}

\title{Towards 6G Networks: Use Cases and Technologies}

\author{{Marco Giordani},~\IEEEmembership{Member, IEEE}, {{Michele Polese},~\IEEEmembership{Member, IEEE},\\{Marco Mezzavilla},~\IEEEmembership{Senior Member, IEEE},
 {Sundeep Rangan},~\IEEEmembership{Fellow, IEEE}, {Michele Zorzi},~\IEEEmembership{Fellow, IEEE}}

\thanks{Marco Giordani, Michele Polese and Michele Zorzi are with the Department of Information Engineering, University of Padova, Padova, Italy (email: \{giordani, polesemi, zorzi\}@dei.unipd.it). Marco Giordani and Michele Polese are primary co-authors.}
\thanks{Marco Mezzavilla and Sundeep Rangan are with NYU WIRELESS, Tandon School of Engineering, New York University, Brooklyn, NY, USA (email: \{mezzavilla, srangan\}@nyu.edu)}}

\makeatletter
\patchcmd{\@maketitle}
  {\addvspace{0.0\baselineskip}\egroup}
  {\addvspace{0\baselineskip}\egroup}
  {}
  {}
\makeatother

\maketitle

\glsunset{nr}

\begin{abstract}
Reliable data connectivity is vital for the ever increasingly intelligent, automated and ubiquitous digital world. 
Mobile networks are the data highways and, in 
a fully connected, intelligent digital world, will need to connect everything, from
people to vehicles, sensors, data, cloud resources and even robotic agents.  
Fifth generation (5G) wireless networks (that are being currently deployed) offer significant advances beyond LTE, but may be unable to meet the full connectivity demands of the future digital society. 
Therefore, this article discusses technologies that will evolve wireless networks towards a sixth generation (6G), and that we consider as enablers for several potential 6G use cases. We provide a full-stack, system-level perspective on 6G scenarios and requirements, and select 6G technologies that can satisfy them either by improving the 5G design, or by introducing completely new communication paradigms.

\end{abstract}

\vspace{-.3cm}

%

\section{Introduction} 

\begin{figure*}[t!]
\centering
\setlength\belowcaptionskip{-.3cm}
\includegraphics[width=.99\textwidth]{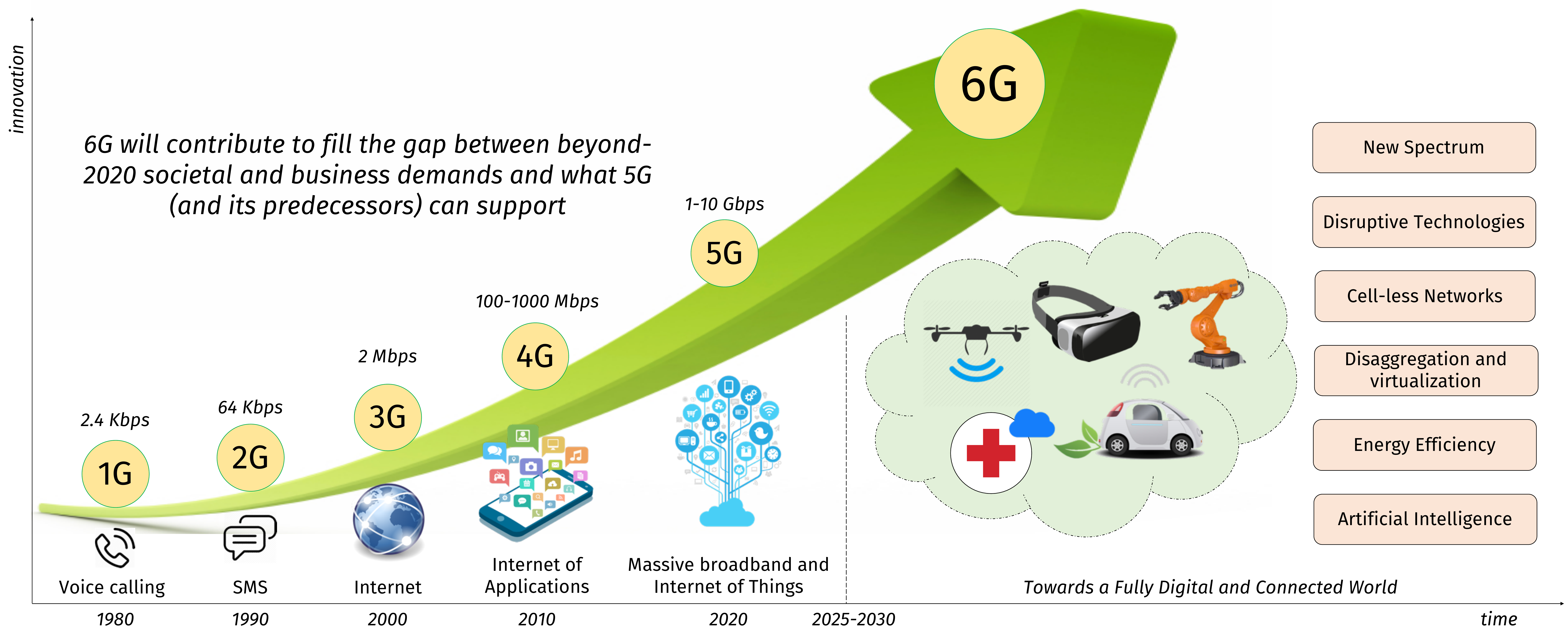}
\caption{Evolution of cellular networks, from 1G to 6G, with a representative application for each generation.}
\label{fig:6ggoals}

\end{figure*}

Each generation of
mobile technology, from the first to the fifth (5G), has been designed to meet the
needs of end users and network operators, as shown in Fig.~\ref{fig:6ggoals}. 
However, nowadays societies are becoming more and more data-centric, data-dependent
and automated.
Radical automation of industrial manufacturing processes will
drive productivity. Autonomous systems are hitting our roads,
oceans and air space. Millions of sensors will be embedded
into cities, homes and production environments, and
new systems operated by artificial intelligence residing in local 'cloud' and 'fog' environments
will enable a plethora of new applications.

Communication
networks will provide the nervous system of these new smart
system paradigms. 
The demands, however, will be daunting.
Networks will need to transfer much greater amounts of data,
at much higher speeds.
Furthering a trend already started in 4G and 5G, sixth generation (6G) connections will move beyond 
personalized communication towards the full realization of the \gls{iot} paradigm, 
connecting not just 
people, but also computing resources, vehicles, devices, wearables, sensors
and even robotic agents~\cite{zhang20196g}. 

5G made a significant step towards developing
a low latency tactile access network, by providing
new additional wireless nerve tracts through (i) new frequency bands (e.g., the \gls{mmwave} spectrum), (ii) advanced spectrum usage and management, in licensed and unlicensed bands, and (iii) a complete redesign of the core network.
Yet, the rapid development of data-centric and automated processes, which require a datarate in the order of terabits per second, a latency of hundreds of microseconds, and $10^7$ connections per km$^2$, may exceed even the capabilities of the emerging 5G systems.

The above discussion has recently motivated researchers to look into a new generation of wireless networks, i.e., sixth generation (6G) systems, to meet
the demands for a fully connected, intelligent digital world.
Along these lines, the broad purpose of this paper is to understand which technologies can identify 6G networks and provide more capable and vertical-specific wireless networking solutions.
Specifically, the paper considers several potential scenarios
for future connected systems, and attempts to estimate their
key requirements in terms
of throughput, latency, connectivity and other factors.  
Importantly, we identify several use cases that go beyond the performance of
the 5G systems under development today, and demonstrate why it is important to think about the long term evolutions beyond 5G. 
Our analysis suggests that, in order 
to meet these demands, radically new communication technologies,
network architectures, and deployment models will be needed.
In particular, we~envision:
\begin{itemize}
	\item \emph{Novel disruptive communication technologies:} although 5G networks have already been designed to operate at extremely high frequencies, e.g., in the mmWave bands in NR, 6G could very much benefit from even higher spectrum technologies, e.g., through Terahertz and optical~communications.
	\item \emph{Innovative network architectures:} despite 5G advancements towards more efficient network setups, the heterogeneity of future network applications and the need for 3D coverage calls for new cell-less architectural paradigms, based on the tight integration of different communication technologies, for both access and backhaul, and on the disaggregation and virtualization of the networking equipment.
	\item \emph{Integrating Intelligence in the Network:} we expect 6G to bring intelligence from centralized computing facilities to end terminals, thereby providing concrete implementation to distributed learning models that have been studied from a theoretical point of view in a 5G context. Unsupervised learning and  knowledge sharing will promote real-time network decisions through~prediction.
\end{itemize}

Prior publications (most notably~\cite{saad2019vision,calvanese2019frontier}) have discussed 6G communications. This article, distinctively, adopts a systematic approach in analyzing the research challenges associated to 6G networks, providing a full-stack perspective, with considerations related to spectrum usage, physical, medium access and higher layers, and network architectures and intelligence for 6G. 
We transfer into our work a multifaceted critical spirit too, having selected, out of several possible innovations,
the solutions that in our view show the highest potential for
future 6G systems. While some of them appear to be incremental, we believe that the combination of breakthrough technologies and evolutions of current networks deserves to be identified as a new generation of mobile networks, as these solutions have not been thoroughly addressed or cannot be properly included in current 5G standards developments, and, therefore, will not be part of commercial 5G deployments. 
We expect our investigation  to promote research efforts towards the definition of new communication and networking technologies to meet the boldest requirements of 6G use cases.

\section{6G Use Cases} 
\label{sec:6g_potential_applications}



5G presents trade-offs on latency, energy, costs, hardware complexity, throughput, and end-to-end reliability. For example, the requirements of mobile broadband and ultra-reliable, low-latency communications are addressed by different configurations of 5G networks. 
6G, on the contrary, will  be developed  to jointly meet stringent network demands (e.g., ultra-high reliability, capacity, efficiency, and low latency) in a holistic fashion, in view of the foreseen economic, social, technological, and environmental context of the 2030 era.

\begin{figure*}[t!]
\centering
	\includegraphics[width=.99\textwidth]{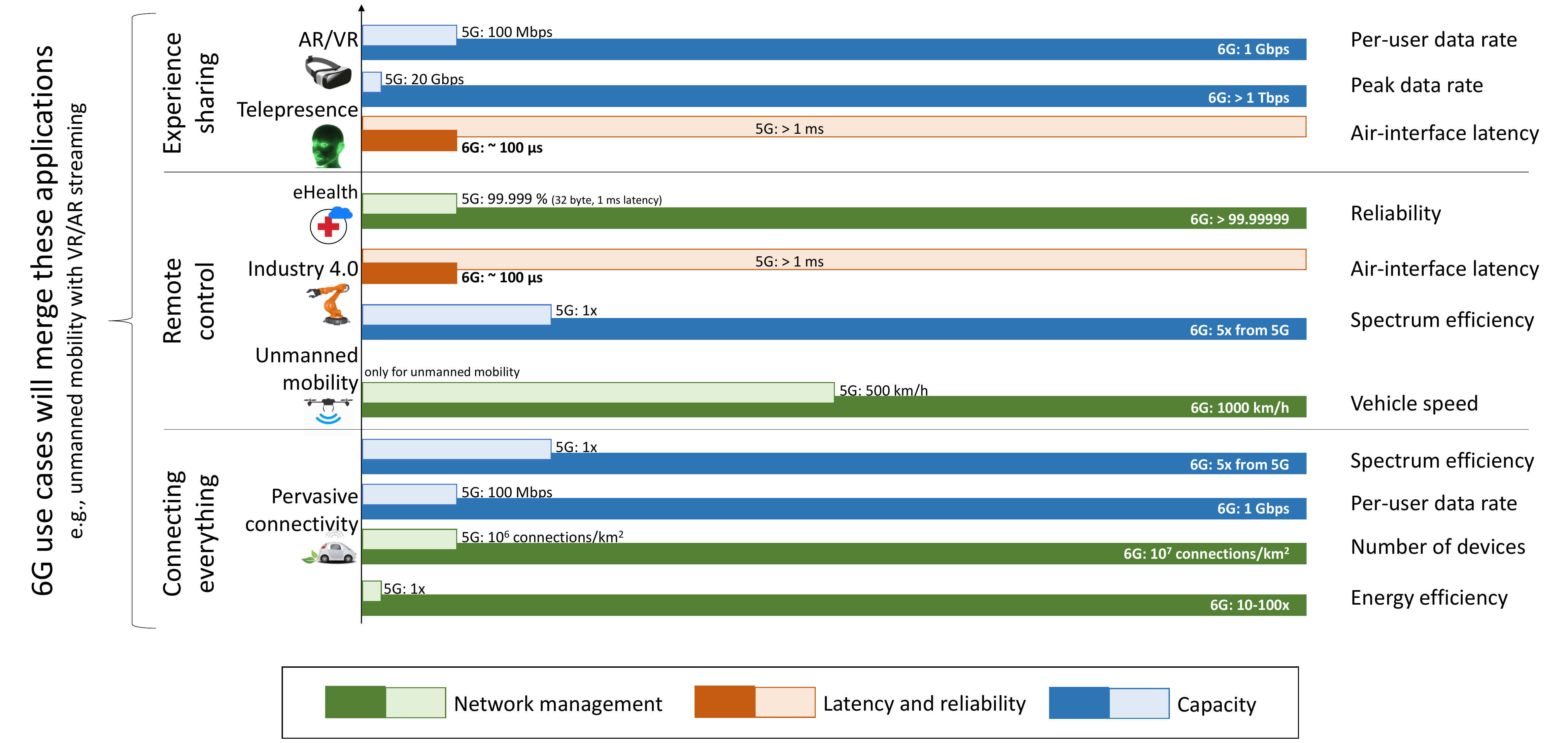}
\caption{Representation of multiple KPIs of 6G use cases, together with the improvements with respect to 5G networks, using data from~\cite{zhang20196g,calvanese2019frontier,saad2019vision,holo1,zhang2018towards,lee2015cyber,wollschlaeger2017future,lu2014connected,choi2016millimeter}.}
\label{fig:kiviat}

\end{figure*}

In this section, we review the characteristics and foreseen requirements of use cases that, for their generality and complementarity, are believed to well represent future 6G services. 
Fig.~\ref{fig:kiviat} provides a comprehensive view on the scenarios in terms of different~\glspl{kpi}.

	\paragraph*{\textbf{Augmented Reality (AR) and Virtual Reality (VR)}} 4G systems unlocked the potential of video-over-wireless, one of the most data-hungry applications at the time. The increasing use of streaming and multimedia services currently justifies the adoption of new spectrum (i.e., mmWaves) to guarantee higher capacity in 5G. However, this multi-Gbps opportunity is attracting new applications which are more data-heavy than bi-dimensional multimedia content: 5G will trigger the \emph{early adoption} of AR/VR. Then, just like video-over-wireless saturated 4G networks, the proliferation of AR/VR applications will deplete the 5G spectrum, and require a system capacity above 1 Tbps, as opposed to the 20 Gbps target defined for 5G~\cite{zhang20196g}. Additionally, to meet the latency requirements that enable real-time user interaction in the immersive environment, AR/VR cannot be compressed (coding and decoding is a time-consuming process), thus the per-user data rate needs to touch the Gbps, in contrast to the more relaxed 100 Mbps 5G target.

	\paragraph*{\textbf{Holographic Telepresence (Teleportation)}} The human tendency to connect remotely with increasing fidelity will pose severe communication challenges in 6G networks.~\cite{holo1} details the datarate requirements of a 3D holographic display: a raw hologram, without any compression, with colors, full parallax, and 30 fps, would require 4.32 Tbps. The latency requirement will hit the sub-ms, and thousands of synchronized view angles will be necessary, as opposed to the few required for VR/AR. Moreover, to fully realize an immersive remote experience, all the 5 human senses are destined to be digitized and transferred across future networks, increasing the overall target data rate. 

	\paragraph*{\textbf{eHealth}} 6G will revolutionize the health-care sector, eliminating time and space barriers through remote surgery and guaranteeing health-care workflow optimizations.
	Besides the high cost, the current major limitation is the lack of real-time tactile feedback~\cite{zhang2018towards}. 
	Moreover, the proliferation of eHealth services will challenge the ability to meet their stringent \gls{qos} requirements, i.e., continuous connection availability (99.99999\% reliability), ultra-low latency (sub-ms), and mobility support.
	The increased spectrum availability, combined with the refined  intelligence of 6G networks, will guarantee these \glspl{kpi}, together with 5-10x gains in spectral efficiency~\cite{zhang20196g}.

	\paragraph*{\textbf{Pervasive Connectivity}} Mobile traffic is expected to grow 3-fold from 2016 to 2021, pushing the number of mobile devices to the extreme, with $10^7$ devices per km$^2$ in dense areas (up from $10^6$ in 5G)~\cite{zhang20196g} and more than 125 billion devices worldwide by 2030. 6G will connect personal devices, sensors (to implement the smart city paradigm), vehicles, and so on. This will stress already congested networks, which will not provide connectivity to every device while meeting the requirements of Fig.~\ref{fig:kiviat}.
	Moreover, 6G networks will require a higher overall energy efficiency (10-100x with respect to 5G), to enable scalable, low-cost deployments, with low environmental impact, and better coverage. Indeed, while 80\% of the mobile traffic is generated indoors, 5G cellular networks,  which are being mainly deployed outdoors and may be operating in the \gls{mmwave} spectrum, will hardly provide indoor connectivity as high-frequency radio signals cannot easily penetrate dielectric materials (e.g., concrete). 
	6G networks will instead provide seamless and pervasive connectivity in a variety of different contexts, matching stringent \gls{qos} requirements in outdoor and indoor scenarios with a cost-aware and resilient infrastructure.

	\paragraph*{\textbf{Industry 4.0 and Robotics}} 6G will fully realize the Industry 4.0 revolution started with 5G, i.e., the digital transformation of manufacturing through cyber physical systems and \gls{iot} services.
	Overcoming the boundaries between the real factory and the cyber
	computational space will enable Internet-based diagnostics, maintenance, operation, and direct machine communications in a cost-effective, flexible and efficient way~\cite{lee2015cyber}. 
	Automation comes with its own set of requirements in terms of reliable and isochronous communication~\cite{wollschlaeger2017future}, which 6G is positioned to address through the disruptive set of technologies we will describe in Sec.~\ref{sec:6g_enabling_technologies}. For example, industrial control requires real-time operations with guaranteed $\mu s$ delay jitter, and Gbps peak data rates for AR/VR industrial applications (e.g., for training, inspection).

	\paragraph*{\textbf{Unmanned mobility}}
	The evolution towards fully autonomous transportation systems offers safer traveling, improved traffic management, and support for infotainment, with a market of 7 trillion~USD~\cite{lu2014connected}.
	Connecting autonomous vehicles demands unprecedented levels of reliability and low latency (i.e., above 99.99999\% and below 1 ms, respectively), even in ultra-high mobility scenarios (up to 1000 km/h), to guarantee passenger safety, a  requirement that is hard to satisfy with existing technologies.
	Moreover, the increasing number of sensors per vehicle will demand increasing data rates (with Terabytes generated per driving hour~\cite{choi2016millimeter}), beyond current network capacity.
	In addition, flying vehicles (e.g., drones) represent a huge potential for various scenarios (e.g., construction, first responders). Swarms of drones will need improved capacity for expanding Internet connectivity. In this perspective, 6G will pave the way for connected vehicles through advances in hardware, software, and the new connectivity solutions we will discuss in Sec.~\ref{sec:6g_enabling_technologies}. 

This wide diversity of use cases is a unique characteristic of the 6G paradigm, whose potential will be fully unleashed only through  breakthrough technological advancements and novel network designs, as described in the next section.


\section{6G Enabling Technologies} 
\label{sec:6g_enabling_technologies}

\begin{figure*}
\centering
\captionof{table}{Comparison of 6G enabling technologies and relevant use cases.}
\includegraphics[width=0.99\textwidth]{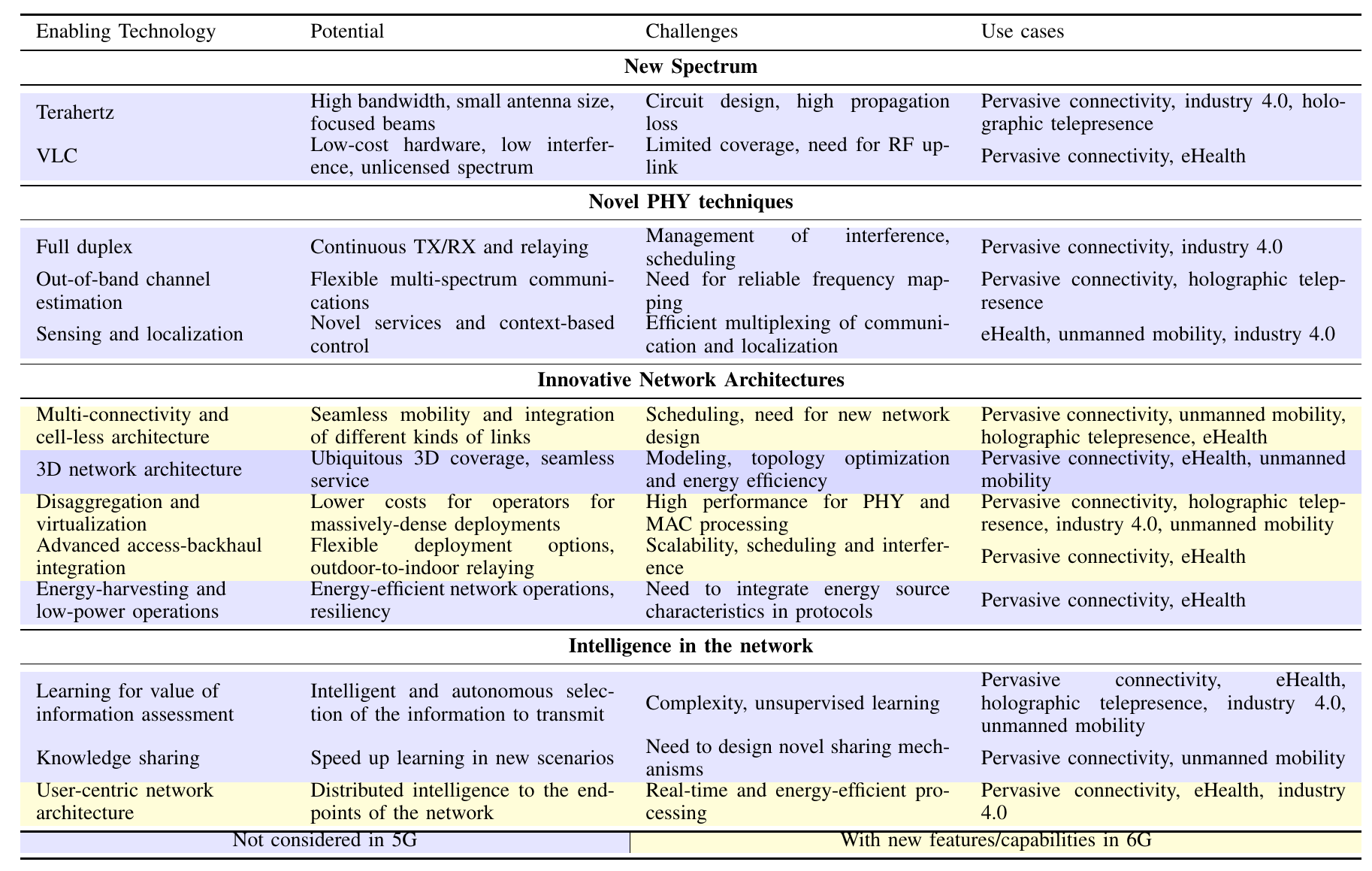}	
\label{table:sumup}
\end{figure*}

In this section, we present the technologies that are rapidly emerging as enablers of the KPIs for the 6G scenarios foreseen in Sec.~\ref{sec:6g_potential_applications}. In particular, Table~\ref{table:sumup}  summarizes potentials and challenges of each proposed technological innovation and suggests which of the use cases introduced in Sec.~\ref{sec:6g_potential_applications} they empower.
Although some of these innovations have already been discussed in the context of 5G,  they were deliberately left out of early 5G standards developments (i.e., 3GPP NR Releases 15 and 16)  and will likely not be  implemented in commercial 5G deployments because of technological limitations or because  markets are not mature enough to support them.
We  consider physical layer breakthroughs in Sec.~\ref{sub:communication_technologies}, new architectural and protocol solutions in Sec.~\ref{sub:innovative_network_architectures}, and finally disruptive applications of artificial intelligence in Sec.~\ref{sub:tools}.

\subsection{Disruptive Communication Technologies} 
\label{sub:communication_technologies}

\begin{figure*}[t!]
\centering
\includegraphics[width=.99\textwidth]{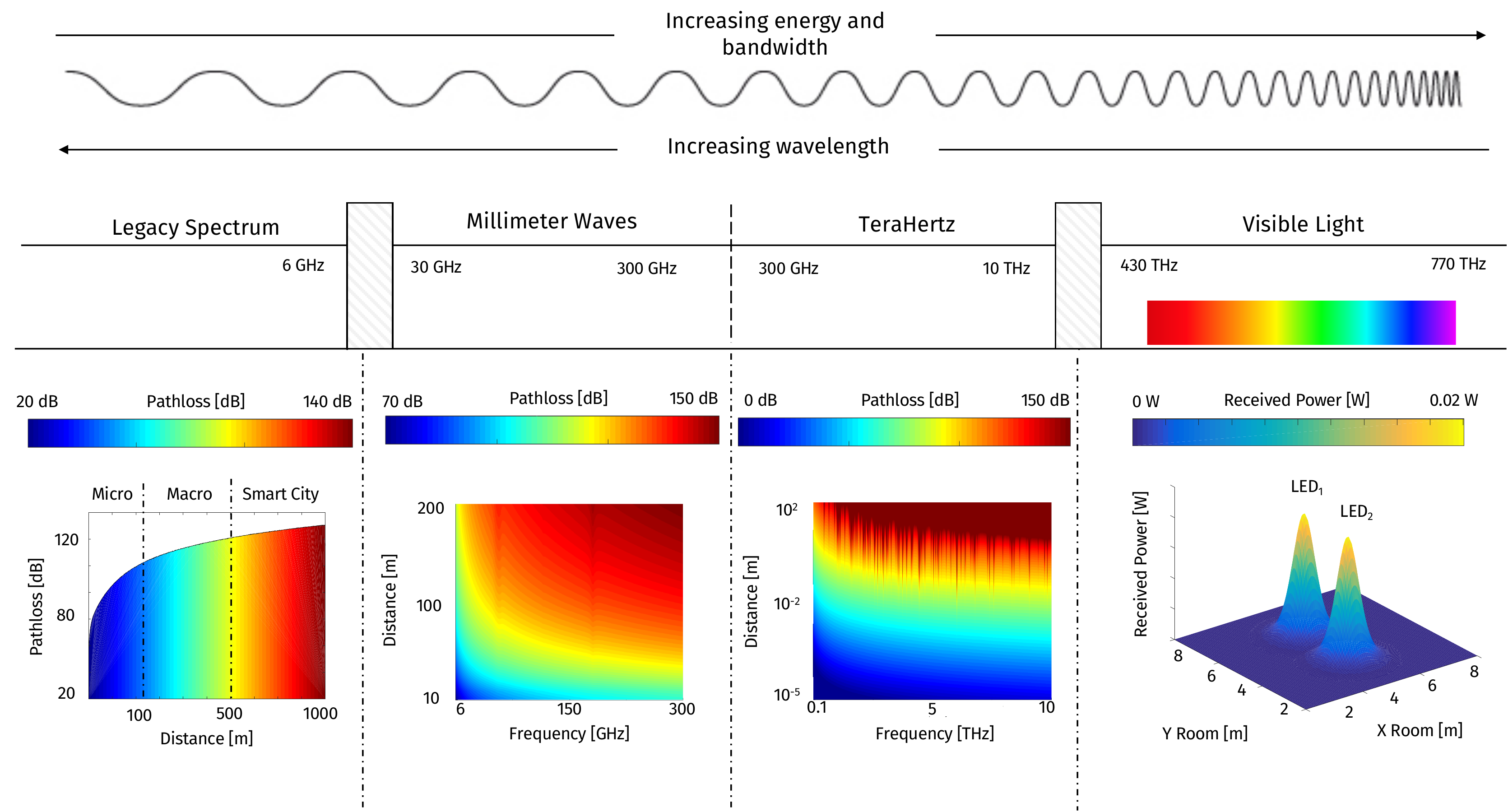}
\caption{Pathloss for sub-6 GHz, mmWave and Terahertz bands, and received power for \gls{vlc}. 
The sub-6 GHz and mmWave pathloss follows the 3GPP models  considering both \gls{los} and \gls{nlos} conditions, while \gls{los}-only is considered for Terahertz~\cite{jornet2011thz} and \gls{vlc}~\cite{komine2004fundamental}.}
\label{fig:pathloss}
\end{figure*}

A new generation of mobile networks is generally characterized by a set of novel communication technologies that provide unprecedented performance (e.g., in terms of available data rate and latency) and capabilities. For example, massive \gls{mimo} and mmWave communications are both key enablers of 5G networks. In order to meet the requirements that we described in Sec.~\ref{sec:6g_potential_applications}, 6G networks are expected to rely on conventional spectrum (i.e., sub-6 GHz and mmWaves) but also on frequency bands that have not yet been considered for cellular standards, namely the Terahertz band and \gls{vlc}. Fig.~\ref{fig:pathloss} represents the pathloss for each of these bands, in typical deployment scenarios, in order to highlight the differences and the opportunities that each portion of the spectrum can exploit. In the following paragraphs, we will focus on the two novel spectrum bands that will be used in 6G.

\begin{itemize}
	\item \textbf{Terahertz communications} operate between 100 GHz and 10 THz~\cite{jornet2011thz} and, compared to mmWaves, bring to the extreme the potential of high-frequency connectivity, enabling data rates in the order of hundreds of Gbps, in line with the boldest 6G requirements. On the other side, the main issues that prevented the adoption of Terahertz links in commercial systems so far are propagation loss, molecular absorption, high penetration loss, and engineering challenges for antennas and \gls{rf} circuitry. As for mmWaves, the propagation loss can be compensated using directional antenna arrays, also enabling spatial multiplexing with limited interference. Furthermore, Terahertz communication performance can be maximized by operating in frequency bands not severely affected by molecular absorption~\cite{jornet2011thz}, as shown in Fig.~\ref{fig:pathloss}. Finally, such high frequencies, when limited to indoor-to-indoor scenarios, enable new kinds of ultra-small-scale electronic packaging solutions for the RF and antenna circuitry.

	\item \textbf{\gls{vlc}} have been proposed to complement \gls{rf} communications by piggybacking on the wide adoption of cheap \gls{led} luminaries. These devices can indeed quickly switch between different light intensities to modulate a signal which can be transmitted to a proper receiver~\cite{pathak2015visible}. The research on \gls{vlc} is more mature than that on Terahertz communications, also thanks to a lower cost of experimental platforms. 
	As reported in Fig.~\ref{fig:pathloss}, \gls{vlc} have limited coverage range, require an illumination source, and suffer from shot noise from other light sources (e.g., the sun), thus can be mostly used indoors~\cite{pathak2015visible}. Moreover, they need to be complemented by \gls{rf} for the uplink. Nonetheless, \gls{vlc} could be used to introduce cellular coverage in indoor scenarios, which, as mentioned in Sec.~\ref{sec:6g_potential_applications}, is a use case that has not been properly addressed by cellular standards. 
\end{itemize}

Although standardization bodies are promoting study items that are oriented towards the investigation of THz and \gls{vlc} solutions for future wireless systems (i.e., IEEE 802.15.3d and 802.15.7, respectively), these technologies have not yet been included in a cellular network standard, and will be targeting beyond 5G use cases.
Moreover, additional research is still required to enable 6G mobile users to operate in the THz and \gls{vlc} spectra, including hardware and algorithms for flexible multi-beam acquisition and tracking in \gls{nlos} environments.

Besides the new spectrum, 6G will also transform wireless networks by leveraging a set of technologies that have been enabled by recent physical layer and circuits research, but are not part of 5G. The following will be key enablers for 6G:
\begin{itemize}
	\item \textbf{Full-duplex communication stack.} With full-duplex communications, the transceivers will be capable of receiving a signal while also transmitting, thanks to carefully designed self-interference-suppression circuits~\cite{goyal2015full}. Practical full-duplex deployments require innovations in antenna and circuit design to reduce the crosstalk between  transmitter and receiver circuits in a wireless device,   thus they have not been included into current cellular network specifications. Future technology advancements, however, will enable concurrent downlink and uplink transmission, to increase the multiplexing capabilities and the overall system throughput without using additional bandwidth. Nonetheless, 6G networks will need careful planning for the full-duplex procedures and deployments, to avoid interference, as well as  novel resource scheduler designs~\cite{goyal2015full}.

	\item \textbf{Novel channel estimation techniques (e.g., out-of-band estimation and compressed sensing).} Channel estimation for directional communications will be a key component of communications at mmWaves and Terahertz frequencies. However, it is difficult to design efficient procedures for directional communications, considering multiple frequency bands and possibly a very large bandwidth. Therefore, 6G systems will need new channel estimation techniques. For example, out-of-band estimation (e.g., for the angular direction of arrival of the signal) can improve the reactiveness of beam management, by mapping the omnidirectional propagation of sub-6 GHz signals to the channel estimation for mmWave frequencies~\cite{ali2018millimeter}. Similarly, given the sparsity in terms of angular directions of mmWave and Terahertz channels, it is possible to exploit compressive sensing to estimate the channel using a reduced number of samples.

	\item \textbf{Sensing and network-based localization.} The usage of \gls{rf} signals to enable simultaneous localization and mapping has been widely studied,
	but such capabilities have never been deeply integrated with the operations and protocols of cellular networks. 6G networks will exploit a unified interface for localization and communications to (i) improve control operations, which can rely on context information to shape beamforming patterns, reduce interference, and predict handovers; and (ii) offer innovative user services, e.g., for vehicular and eHealth applications.
\end{itemize}

\begin{figure*}[t!]
\centering
\includegraphics[width=.99\textwidth]{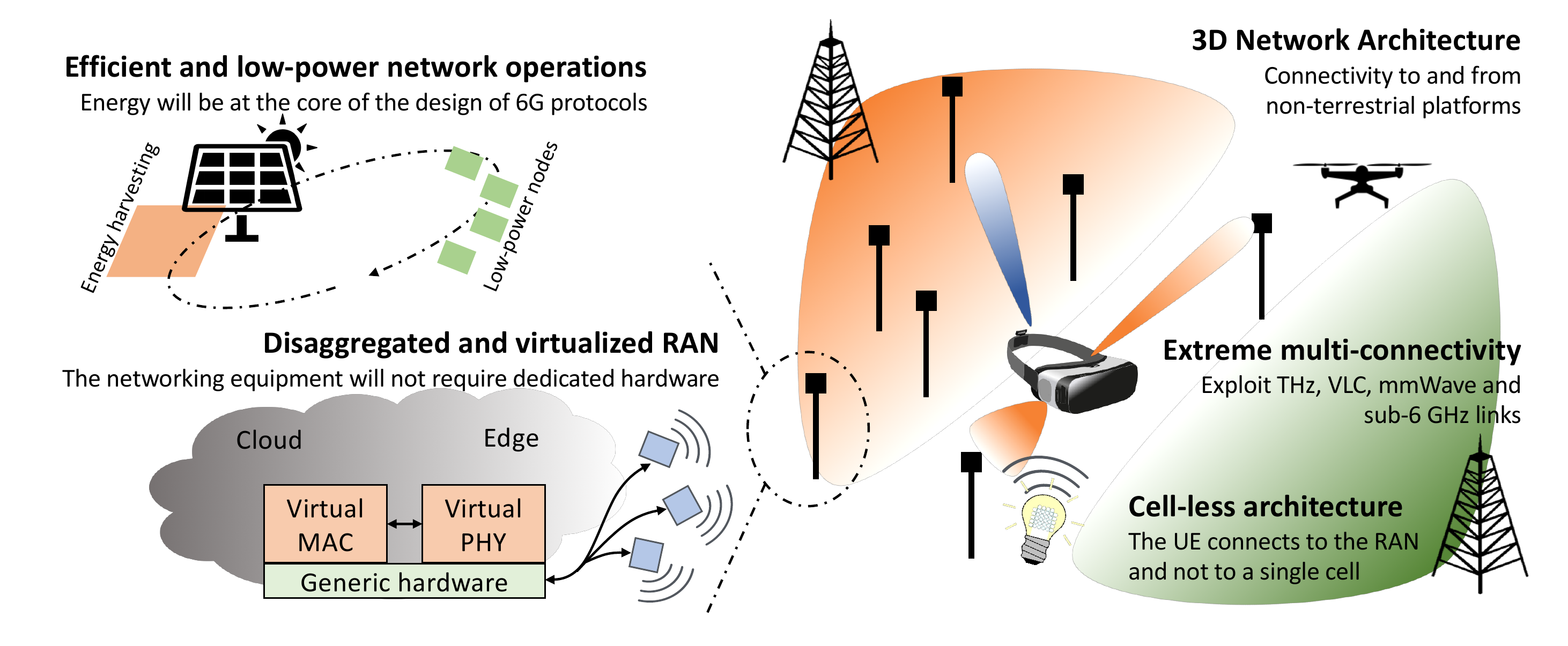}
\caption{Architectural innovations introduced in 6G networks.}
\label{fig:architecture}
\end{figure*}

\subsection{Innovative Network Architectures} 
\label{sub:innovative_network_architectures}

The disruption brought by the communication technologies described in Sec.~\ref{sub:communication_technologies} will enable a new 6G network architecture, but also potentially require structural updates with respect to current mobile network designs. For example, the density and the high access data rate of Terahertz communications will increase the capacity demands on the underlying transport network, which has to provide both more points of access to fiber and a higher capacity than today's backhaul networks. Moreover, the wide range of different communication technologies available will increase the heterogeneity of the network, which will need to be managed.

The main architectural innovations that 6G will introduce are described in Fig.~\ref{fig:architecture}. In this context, we envision the introduction and/or deployment of the following~paradigms: 

\begin{itemize}
	\item \textbf{Tight integration of multiple frequencies and communication technologies and cell-less architecture.} 6G devices will support a number of heterogeneous radios in the devices. This enables multi connectivity techniques that can extend the current boundaries of cells, with users connected to the network as a whole (i.e., through multiple complementary technologies) and not to a single cell. The cell-less network procedures will guarantee a seamless mobility support, without overhead due to handovers (which might be frequent when considering systems at Terahertz frequencies), and will provide \gls{qos} guarantees that are in line with  the most challenging mobility requirements envisioned for 6G, as in the vehicular scenarios. The devices will be able to seamlessly transition among different heterogeneous links (e.g., sub-6 GHz, mmWave, Terahertz or \gls{vlc}) without manual intervention or configuration. Finally, according to the specific use case, the user may also concurrently use different network interfaces to exploit their complementary characteristics, e.g., the sub-6 GHz layer for control, and a Terahertz link for the data plane.

	\item \textbf{3D network architecture.} 5G networks (and previous generations) have been designed to provide connectivity for an essentially bi-dimensional space, i.e., network access points are deployed to offer connectivity to devices on the ground.
	On the contrary, we envision future 6G heterogeneous architectures to provide three-dimensional (3D) coverage, thereby complementing terrestrial infrastructures with non-terrestrial platforms (e.g., drones, balloons, and satellites). Moreover, these elements could also be quickly deployed to guarantee seamless service continuity and reliability, e.g., in rural areas or during events, avoiding the operational and management costs of always-on, fixed infrastructures.
	Despite such promising opportunities, there are various challenges  to be solved before flying platforms can effectively be used in wireless networks, e.g., air-to-ground channel modeling, topology and trajectory optimization, resource management and energy efficiency. 

	\item \textbf{Disaggregation and virtualization of the networking equipment.} Even though networks have recently started to transition towards the disaggregation of once-monolithic networking equipments,  the 3GPP does not directly specify how to introduce virtualization concepts. Moreover, current 5G studies have not yet addressed the challenges related to the design of disaggregated architectures that can operate under the higher control latency that might be introduced by centralization, and to the security of virtualized network functions, which could be subjected to cyber-attacks.
  	6G networks will bring  disaggregation to the extreme by virtualizing  \gls{mac} and \gls{phy} layer components which currently require dedicated hardware implementations, and realizing low-cost distributed platforms with just the antennas and minimal processing. This will decrease the cost of networking equipment, making a massively dense deployment economically feasible.

	\item \textbf{Advanced access-backhaul integration.} The massive data rates of the new 6G access technologies will require an adequate growth of the backhaul capacity. Moreover, Terahertz and \gls{vlc} deployments will increase the density of access points, which need backhaul connectivity to their neighbors and the core network. The huge capacity of 6G technologies can thus be exploited for self-backhauling solutions, where the radios in the base stations provide both access and backhaul. While a similar option is already being considered for 5G, the scale of 6G deployments will introduce new challenges and opportunities, e.g., as the networks will need higher autonomous configuration capabilities.

	\item \textbf{Energy-harvesting strategies for low power consumption network operations.} 
	Incorporating energy-harvesting mechanisms into 5G infrastructures currently faces several issues, including coexistence with the communications, and efficiency loss when converting harvested signals to electric current. Given the scale expected in 6G networks, it is necessary to design  systems where both the circuitry and the communication stack are developed with energy-awareness in mind. One option is using energy harvesting circuits to allow devices to be self-powered, which could be critical for example to enable off-grid operations, long-lasting \gls{iot} devices and sensors, or long stand-by intervals for devices and equipment which are rarely used.


\end{itemize}

\subsection{Integrating Intelligence in the Network} 
\label{sub:tools}

The complexity of 6G communication technologies and network deployments will probably prevent closed-form and/or manual optimizations. While intelligent techniques in cellular networks are already being discussed for 5G, we expect 6G deployments to be much denser (i.e., in terms of number of access points and users), more heterogeneous (in terms of integration of different technologies and application characteristics), and with stricter performance requirements with respect to 5G. Therefore,  intelligence will play a more prominent role in the network, going beyond the classification and prediction tasks which are being considered for 5G systems. Notice that the standard may not specify the techniques and learning strategies to be deployed in networks, but data-driven approaches can be seen as tools that network vendors and operators can use to meet the 6G requirements~\cite{wang2018machine}.  In particular,  6G research will be oriented towards the following aspects:
\begin{itemize}
\item \textbf{Learning techniques for data selection and feature extraction.}
The large volume of data generated by future connected devices (e.g., sensors in autonomous vehicles) will put a strain on communication technologies, which could not guarantee the required quality of service. It is therefore fundamental to discriminate the \emph{value of information} to maximize the utility for the end users with (limited) network resources.
In this context, \gls{ml} strategies can evaluate the degree of correlation in observations, or extract features from input vectors and predict the a-posteriori probability of a sequence given its entire history.
In 6G, unsupervised and reinforcement learning approaches, moreover, do not need labeling and can be used to operate the network in a truly autonomous fashion.

\item \textbf{Inter-user inter-operator knowledge sharing.}
Spectrum and infrastructure sharing is beneficial in cellular networks, to maximize the multiplexing capabilities. With learning-driven networks, operators and users can also share learned/processed representations of specific network deployments and/or use cases, for example to speed up the network configuration in new markets, or to better adapt to new unexpected operational scenarios. 
The trade-offs in latency, power consumption, system overhead, and cost will be studied in 6G, for both on-board and edge-cloud-assisted solutions.

\item \textbf{User-centric network architecture.} \gls{ml}-driven networks are still in their infancy, but will be a fundamental component of complex 6G systems, which envision distributed artificial intelligence, to implement a fully-user-centric network architecture.
In this way, end terminals will be able to make autonomous network decisions based on the outcomes of previous operations, without communication overhead to and from centralized controllers. 
Distributed methods can process \gls{ml} algorithms in real time, i.e., with a sub-ms latency, as required by several 6G~services, thereby yielding more responsive network management.

\end{itemize}

\section{Conclusions} 
\label{sec:conclusions}

In this paper, we reviewed  use cases and  technologies that we believe will characterize 6G networks. Table~\ref{table:sumup} summarizes the main challenges, potentials and use cases of each enabling technology. 6G wireless research can disrupt the traditional cellular networking paradigms that still exist in 5G, introducing for example the support for Terahertz and visible light spectra, cell-less and aerial architectures, and massive distributed intelligence, among others. These technologies, however, are not market-ready: this represents a unique opportunity for the wireless research community to foster innovations that will enable unforeseen digital use cases for the  society of 2030 and~beyond. 

\section*{Acknowledgements}
This work was partially supported by NIST through Award No. 70NANB17H166, by the U.S. ARO under Grant no. W911NF1910232, by MIUR (Italian Ministry for Education and Research) under the initiative "Departments of Excellence" (Law 232/2016), by NSF grants 1302336, 1564142, and 1547332, the SRC and the industrial affiliates of NYU WIRELESS.

\bibliographystyle{IEEEtran}

\vspace*{-1cm}%
\begin{IEEEbiographynophoto}{Marco Giordani}
[M'20] was a Ph.D. student in Information Engineering at the University of Padova, Italy (2016-2019),  where he is now a postdoctoral researcher and adjunct professor.
He visited  NYU and TOYOTA Infotechnology Center, Inc., USA.
In 2018 he received the “Daniel E. Noble Fellowship Award” from the IEEE Vehicular Technology Society. His research  focuses on protocol design for 5G mmWave cellular and vehicular networks.
\end{IEEEbiographynophoto}%
\vspace*{-1cm}%
\begin{IEEEbiographynophoto}{Michele Polese}
[M'20] was a Ph.D. student in Information Engineering at the University of Padova (2016-2019), where he is now a postdoctoral researcher and adjunct professor. He visited NYU, AT\&T Labs, and Northeastern University. His research focuses on protocols and architectures for 5G mmWave networks.
\end{IEEEbiographynophoto}%
\vspace*{-1cm}%
\begin{IEEEbiographynophoto}{Marco Mezzavilla}
[SM'19] is a research scientist at the NYU Tandon School of Engineering. He received his Ph.D. (2013) in Information Engineering from the University of Padova, Italy. His research focuses on design and validation of communication protocols and applications of 4G/5G technologies.
\end{IEEEbiographynophoto}%
\vspace*{-1cm}%
\begin{IEEEbiographynophoto}{Sundeep Rangan}
[F'15] is an ECE professor at NYU and Associate 
Director of NYU WIRELESS. He received his Ph.D. from the University of California, Berkeley. In 2000, he co-founded (with four others) Flarion Technologies, a spinoff of Bell Labs that developed the first cellular OFDM data system. It was acquired by Qualcomm in 2006, where he was a director of engineering prior to joining NYU in 2010.
\end{IEEEbiographynophoto}%
\vspace*{-1cm}%
\begin{IEEEbiographynophoto}{Michele Zorzi}
[F'07] is with the Information Engineering Department of the University of Padova, focusing on wireless communications research. He was Editor-in-Chief of IEEE Wireless Communications from 2003 to 2005, IEEE Transactions on Communications from 2008 to 2011, and IEEE Transactions on Cognitive Communications and Networking from 2014 to 2018. He served ComSoc as a Member-at-Large of the Board of Governors from 2009 to 2011, as Director of Education and Training from 2014 to 2015, and as Director of Journals from 2020 to 2021.
\end{IEEEbiographynophoto}

\end{document}